\documentclass[prb,aps,superscriptaddress,amsmath,amssymb,floatfix,twocolumn,showpacs,amsfonts,longbibliography]{revtex4-2}
\usepackage{times}
\usepackage[varg]{txfonts}
\usepackage{threeparttable}
\usepackage{textcomp}
\usepackage{graphicx}
\usepackage{subfigure}
\usepackage{subeqnarray}
\usepackage{mathrsfs}
\usepackage{mathtools}
\usepackage{amsmath}
\usepackage{amssymb}
\usepackage{color}
\usepackage[colorlinks=true,citecolor=blue,urlcolor=blue,linkcolor=blue,hyperindex]{hyperref}
\usepackage{braket}
\usepackage{overpic}
\usepackage{ulem}
\usepackage[version=3]{mhchem}

\begin{document}

\title{Landau levels and optical conductivity in the mixed state of a generic Weyl superconductor}

\author{Zhihai Liu}
\affiliation{College of Physics and Optoelectronic Engineering, Shenzhen University, Shenzhen 518060, China}

\author{Luyang Wang}
\email{wangly@szu.edu.cn}
\affiliation{College of Physics and Optoelectronic Engineering, Shenzhen University, Shenzhen 518060, China}

\date{\today}

\begin{abstract}
The low-energy quasiparticle states in the mixed state of most superconductors remain Bloch waves due to the presence of supercurrent around vortex cores. In contrast, the Weyl superconductor (WSC) may display Dirac-Landau levels in the presence of a vortex lattice. Here, we investigate the Landau level (LL) structure and optical conductivity in the mixed state of a generic WSC using a heterostructure model, where the tilt of the Bogoliubov-Weyl (BW) cones can be tuned, yielding either type-I (undertilted) or type-II (overtilted) cones. We find that, in a magnetic field, the tilted type-I BW cone in the mixed state may exhibit squeezed LLs with reduced spacings. On the other hand, the spectrum of type-II cones shows a dependence on the angle between the magnetic field and the tilt direction; LL quantization is only possible if the angle is below a critical value. For zero tilt, the optical conductivity in the mixed state of the WSC shows peaks only at photon frequencies $\omega_n \propto \sqrt{n}+\sqrt{n+1}$, with a linear background. However, the tilt of BW cones results in the emergence of optical transitions beyond the usual dipolar ones. For type-II BW cones, unique intraband transition conductivity peaks emerge at low frequency, which can serve as an indicator to distinguish between type-I and type-II BW cones in WSCs.
\end{abstract}

\maketitle

\section{Introduction}\label{sec:intro}
In a strong magnetic field, the electron states of solid materials are typically reorganized into a series of discrete energy levels called Landau levels. These discrete levels can give rise to unique physical phenomena, such as the periodic quantum oscillations of certain quantities~\cite{shoenberg2009,shoenberg2013,LifshitsJPCS1958,PotterNatureC2014} and the quantum Hall effect~\cite{Ezawa2008quantum,GusyninPRL2005,NovoselovNatureP2006,GoerbigRMP2011}. However, such scenarios are uncommon in superconductors. For the first-kind superconductors, the magnetic field is completely repelled out of the sample due to the Meissner effect. For the second-kind superconductors, the magnetic field partially penetrates the sample, resulting in the mixed state, where Abrikosov vortex lattice is formed. In this state, the spatially varying supercurrent disrupts the system's LLs~\cite{ASMelnikovJPCM1999,Yasui.PRL1999,Franz.PRL2000, Marinelli.PRB2000,Kopnin.PRB2000,Morita.PRL2001,Knapp.PRB2001,Vishwanath.PRL2001,Vafek.PRL2006,Melikyan.PRB2007}. Particularly, Franz and Tesanovic~\cite{Franz.PRL2000} have revealed that, in the vortex state of a $d$-wave superconductor, the magnetic field vanishes on average, the quasiparticles retain the zero-field Dirac cone and the main effect of the magnetic field is the renormalization of the Fermi velocity. Similar phenomena have also been reported in $s$-wave and $p$-wave superconductors~\cite{Vafek.PRB63.2001}. It is believed that most superconductors do not exhibit LLs in a magnetic field.

Recently, Pacholski \textit{et al.}~\cite{Pacholski.PRL2018} have demonstrated that due to the chirality of the Weyl fermions, WSCs may exhibit topologically protected zeroth LL in the vortex state. It is also noted that the pseudo-LL structure induced by a pseudomagnetic field have been reported in WSCs and other nodal superconductors~\cite{Massarelli.PRB2017,Tianyu.PRB2017,NicaPRB2018}. The WSC can be considered as a superconducting counterpart of the Weyl semimetal (WSM), which exhibits BW nodes within the superconducting gap~\cite{Meng.PRB2012,Yang.PRL2014,Bednik.PRB2015,Meng.PRB2017,Nakai.PRB2020,SatoJPSJ2016}. Consequently, some physical properties that resemble those found in WSMs have been reported in WSCs. For example, chiral Majorana arcs at zero energy have been examined on the surface Brillouin zone (BZ) of WSCs, analogous to surface Fermi arcs of WSMs~\cite{WanXiangangPRB2011}. These Majorana arcs may lead to the anomalous thermal Hall effect~\cite{Meng.PRB2012,Nakai.PRB2020}. In an electromagnetic field, the WSM shows negative magnetoresistance induced by the chiral anomaly~\cite{SonPRB2013}. Although the charge is not conserved in the superconducting state due to the $U(1)$ symmetry breaking, WSCs may display negative thermal magnetoresistivity since energy is still conserved. In the superconducting state, the chiral anomaly can be induced by vortex textures in the mixed state~\cite{KobayashiPRL2018} or by lattice strain~\cite{Guinea.NatureP2010,Levy.Science2010,ShapourianPRB2015,LiuTianyuPRB95.2017}. WSCs may be realized in some multilayer structures that include $s$-wave superconductors~\cite{Meng.PRB2012,Nakai.PRB2020}. Furthermore, materials such as \ce{URu_2Si_2}~\cite{KasaharaPRL2007,YamashitaNatureP2015}, \ce{UPt_3}~\cite{TouHPRL1998,GoswamiPRB2015}, \ce{UCoGe}~\cite{HuyNTPRL2007,deVisserPRL2009} and \ce{SrPtAs}~\cite{BiswasPKPRB2013,FischerPRB2014} are also supposed to be  the candidates for WSCs.

In WSMs, Weyl cones can be tilted in the momentum-energy space and are classified into type-I and type-II cones based on the ratio between the tilt and Fermi velocity~\cite{BurkovNatureMat2016,SoluyanovNature2015}. Similarly, WSCs may present tilted BW cones, including both type-I and type-II cones\cite{Xu2015PRL}. In two-dimensional (2D) systems, the impact of tilt on the LLs of Dirac cones has been investigated. For instance, in the quasi-2D organic compound \ce{$\alpha$-(BEDT-TTF)_2I_3}, the tilted Dirac cone shows squeezed LLs that collapse when the tilt exceeds the Fermi velocity~\cite{GoerbigEPL2009,JuditPRB2015}. The modification of LLs induced by the tilt can be viewed as being generated by an in-plane electric field, which has been theoretically investigated in the context of graphene by Lukose \textit{et al.}~\cite{LukosePRL2007}. Due to the Lorentz covariance of Dirac fermions, below the critical electric field, we can always find a frame of reference, achieved by a ``Lorentz boost", in which the in-plane electric field effectively vanishes and the magnetic field is reduced, the so-called magnetic regime~\cite{jackson1999}. The LL strcture of tilted Dirac cones can also be interpreted through a generalized chiral symmetry~\cite{KawarabayashiPRB2011,KawarabayashiIJMP2012}. Similar phenomena have been reported in the 3D WSM, where the squeezed LLs persist as long as the tilt in the plane perpendicular to the magnetic field does not exceed the Fermi velocity,  even for type-II WSMs~\cite{TchoumakovPRL2016,YuZhiMingPRL2016}. Otherwise, the LLs collapse, leading the system into the ``electric regime" where the magnetic field effectively vanishes.

In the mixed state of a WSC, the qsuasiparticle spectrum for the untilted BW cones has been studied, revealing Dirac-LLs~\cite{Pacholski.PRL2018}. In the plane perpendicular to the magnetic field, these LLs scale with $\sqrt{nB}$ and feature completely dispersionless zeroth levels. However, the impact of the tilt on the LL structure of BW cones has not been fully elucidated. Moreover, the quasiparticle spectrum in the mixed state for the type-II BW cones remains ambiguous. In this work, we propose a heterostructure model that is engineered by multilayer structures comprising WSMs and $s$-wave superconductors. The heterostructure may present a Weyl superconducting phase due to the superconducting proximity effect~\cite{FuLiangPRL2008,KhannaPRB2014,ChenAnffanyPRB2016,PBaireuther.NJP2017}, the quasiparticle spectrum of which shows tilted type-I and type-II BW nodes in the superconducting gap. In particular, the type-II BW node connects an electron-like and a hole-like Bogoliubov Fermi pocket. We find that the tilt modifies the LLs of the WSC in a similar manner to that of WSMs. Specifically, for the type-II BW cones, when the projection of the tilt in the plane perpendicular to the magnetic field exceeds the Fermi velocity, the LLs collapse, and the quasiparticle states in the vortex lattice become Bloch waves.

Ahn and Nagaosa have shown that intrinsic momentum-conserving optical excitations may occur in clean multi-band superconductors~\cite{Ahn.NatureC2021}. Indeed, optical responses in several clean superconducting systems have been reported~\cite{ZhaoLNatureP2017,XuTianruiPRB2019,KamataniPRB2022,PapajPRB2022}. In contrast, in the single-band superconductors, optical excitations usually require impurity scattering. For example, the ac anomalous Hall conductivity $\sigma_H(\omega)$ induced by impurities in chiral superconductors has been investigated~\cite{HuangWenPRR2020,HuangWenPRB2023}. Here, we study optical responses in the mixed state of the generic WSC, where vortices serve as impurities, and find similar optical characteristics to the magneto-optical conductivity of WSMs~\cite{AshbyPRB2013,TchoumakovPRL2016,YuZhiMingPRL2016,MarcusPRB2020,LiuPuPRB2023}. Specifically, the magneto-optical conductivity for untilted BW cones displays a series of peaks at photon frequencies $\omega_n \propto \sqrt{n}+\sqrt{n+1}$, with a linear background that arises from the dispersion of LLs in the magnetic direction. For a finite tilt in the plane perpendicular to the magnetic field, emerging transitions beyond the usual dipolar selection rule $n\to n \pm 1$ give rise to new peaks in the optical conductivity. In particular, unique intraband transitions emerge in the magneto-optical conductivity of type-II BW cones, leading to peaks at low frequency that cannot be observed for type-I BW cones.

\section{Weyl-Superconductor Model}\label{sec:model}
\begin{figure}[t]
  \centering
  \subfigure{\includegraphics[width=1.6in]{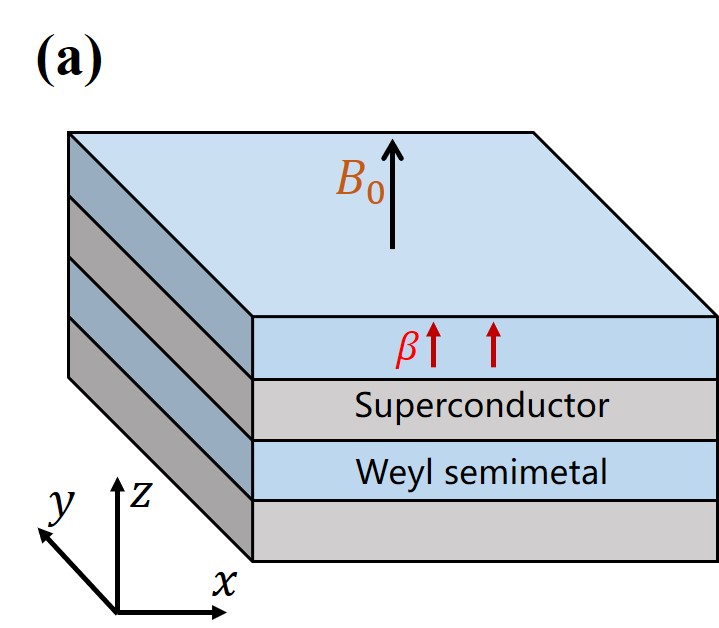}\label{fig:hstrut}}~~
  \subfigure{\includegraphics[width=1.7in]{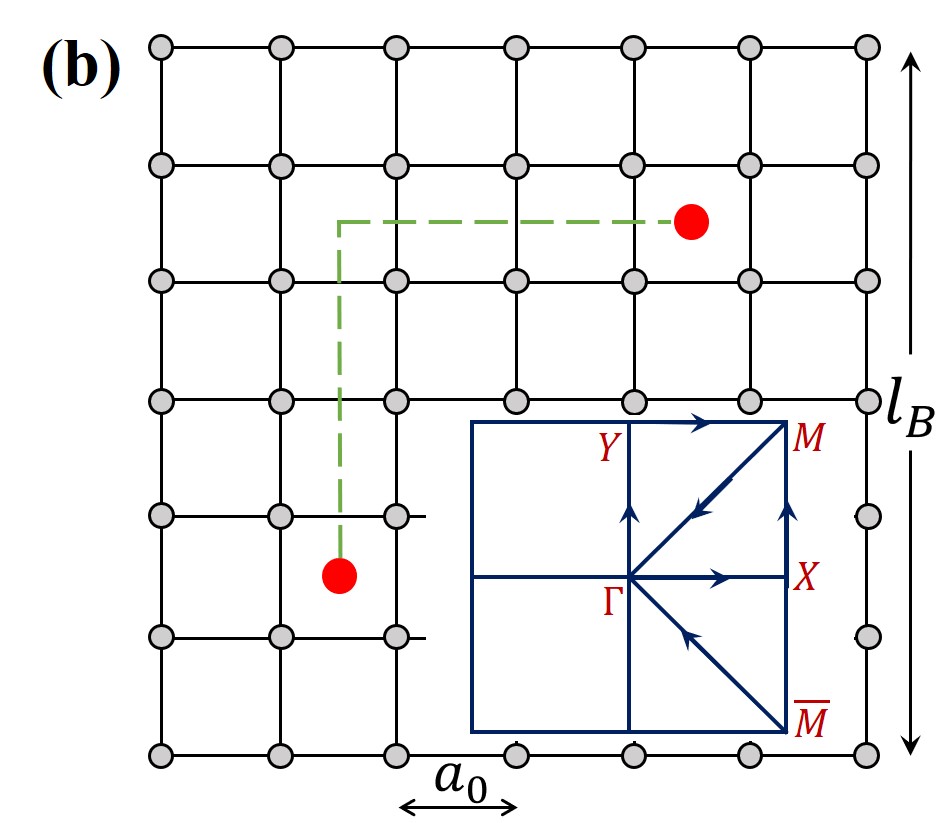}\label{fig:Lattice}}
  \caption{(a) The WSC realized by a WSM-superconductor heterostructure. A magnetic field $B_0$ is applied perpendicularly to the multilayer structure, and the WSM has an intrinsic magnetization $\beta$. (b) The projection of the vortex lattice in the $x$-$y$ plane, where each magnetic unit cell, with a constant $l_B=6a_0$, includes two vortices (red disks). The branch cut connecting two vortices is shown by the green dashed line. The inset in (b) displays the planar BZ in the $k_x$-$k_y$ plane, where the arrows indicate some high symmetry directions.} \label{fig:hslat}
\end{figure}

We start from a heterostructure engineered by alternately stacking WSM and $s$-wave superconductor layers, as shown in Fig.~\ref{fig:hstrut}. For simplicity, we consider a two-band magnetic WSM model in which the time-reversal symmetry is broken while the inversion symmetry is preserved. 
The Hamiltonian
is written as~\cite{ArmitageRMP2018}
\begin{eqnarray}
\begin{aligned}
H_0({\bf k}) &= t_0(\sigma_{x}\sin{k_x a_0}+\sigma_{y}\sin{k_y a_0}) + M({\bf k})\sigma_z - \mu\sigma_0, \\
M({\bf k}) &= t_0 \lambda(\beta - \cos{k_x a_0} - \cos{k_y a_0}) + t_z \cos{k_z a_0},
\end{aligned}
\label{EQ:HamWM}
\end{eqnarray}
where $\sigma_i$'s represent the Pauli matrices, $\mu$ is the chemical potential. The model Eq.~(\ref{EQ:HamWM}) can be realized in a cubic lattice with the same lattice constant $a_0$ (setting $a_0\equiv1$) in the $x$, $y$ and $z$ directions. The magnetization $\beta$ breaks the time-reversal symmetry. Unless otherwise specified, we take $\beta=2$, $\lambda=1$ and $\mu=0$ in this work, setting $t_z=t_0=1$ as the energy unit. The term with $\lambda$ ($\lambda\neq 0$) lifts the degeneracy at points $(0,\pi)$, $(\pi,0)$ and $(\pi,\pi)$ in the $k_z=\pm \tfrac{\pi}{2}$ planes, leaving only two Weyl nodes located at $(0,0,\pm \tfrac{\pi}{2})$, respectively.

In the WSM layers, conventional superconductivity is induced by the superconducting proximity effect, which is governed by a Bogoliubov-de Gennes (BdG) Hamiltonian. This Hamiltonian describes the coupling of electrons and holes by the pair potential and is written as
\begin{eqnarray}
H_{BdG}({\bf k})=
\begin{pmatrix}
H_0({\bf k}) & \Delta \\
\Delta^* & -\sigma_yH_0^*(-{\bf k})\sigma_y
\end{pmatrix},
\label{EQ:HamBDG}
\end{eqnarray}
where the superconducting pair potential $\Delta=\Delta_0e^{i\phi}$, with $\phi$ being the globally coherent superconducting phase.

\begin{figure}[ht]
  \centering
  \subfigure{\includegraphics[width=1.6in]{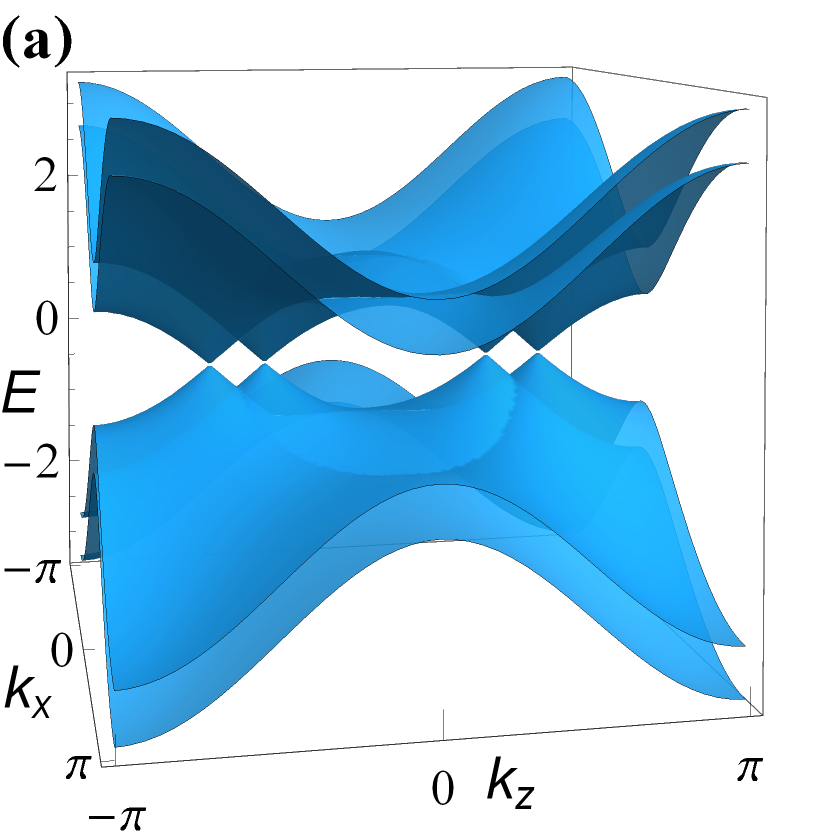}\label{fig:Banda}}~~
  \subfigure{\includegraphics[width=1.6in]{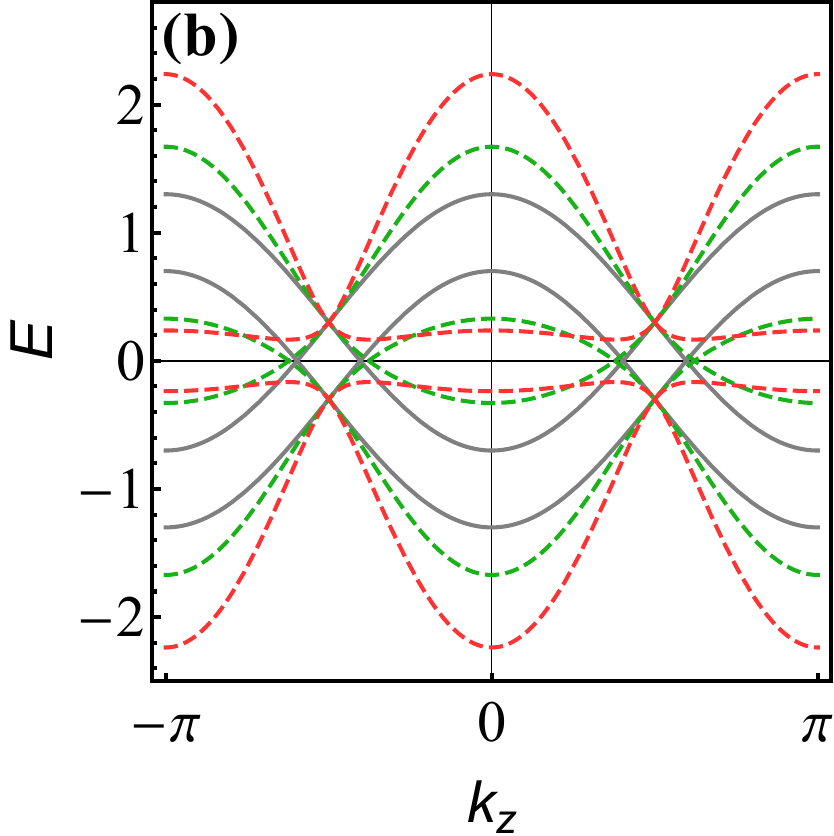}\label{fig:Bandb}}
  \subfigure{\includegraphics[width=1.6in]{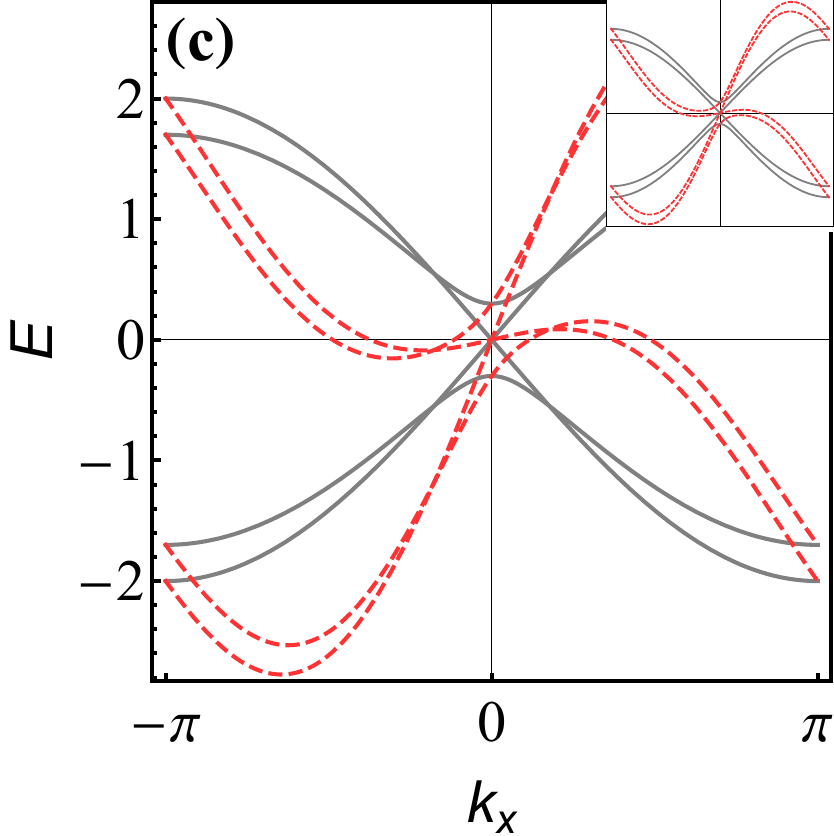}\label{fig:Bandc}}~~
  \subfigure{\includegraphics[width=1.6in]{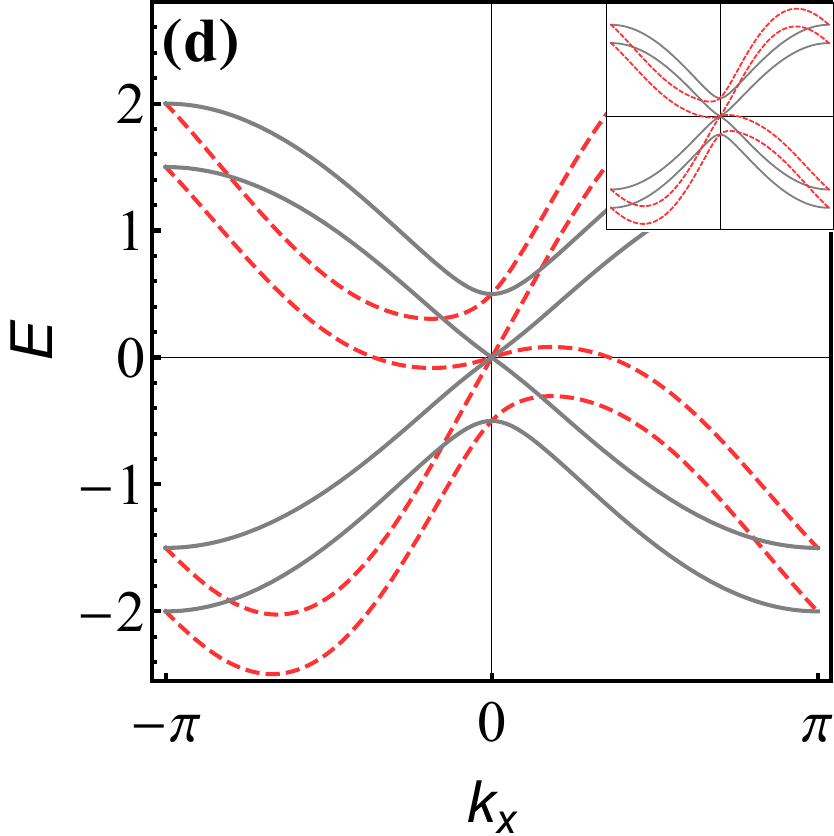}\label{fig:Bandd}}
  \subfigure{\includegraphics[width=1.6in]{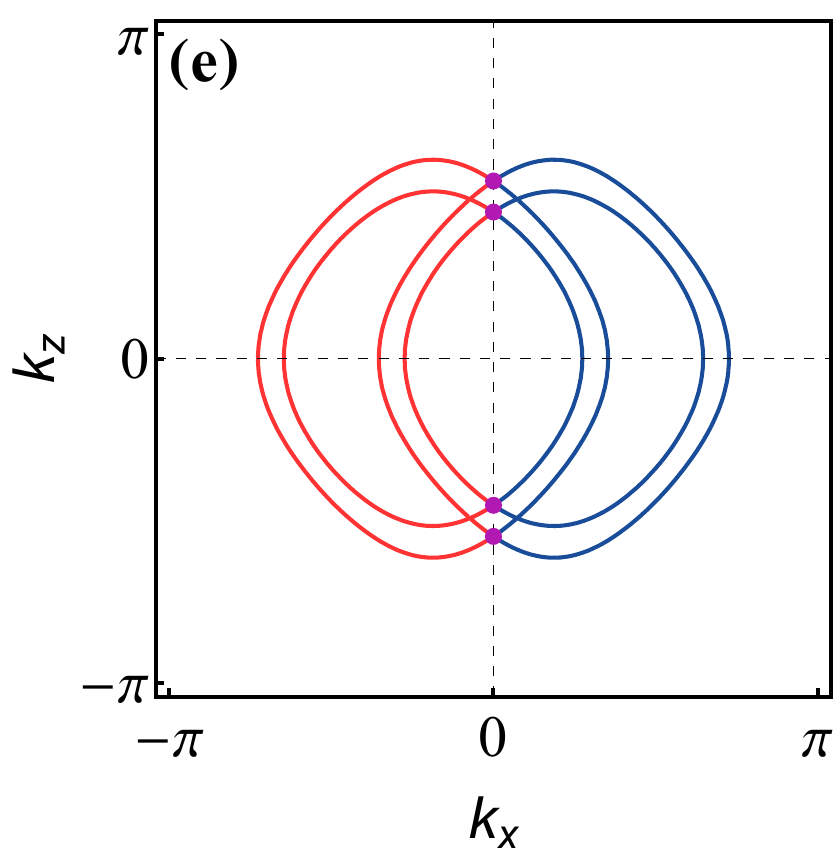}\label{fig:Bande}}~~
  \subfigure{\includegraphics[width=1.6in]{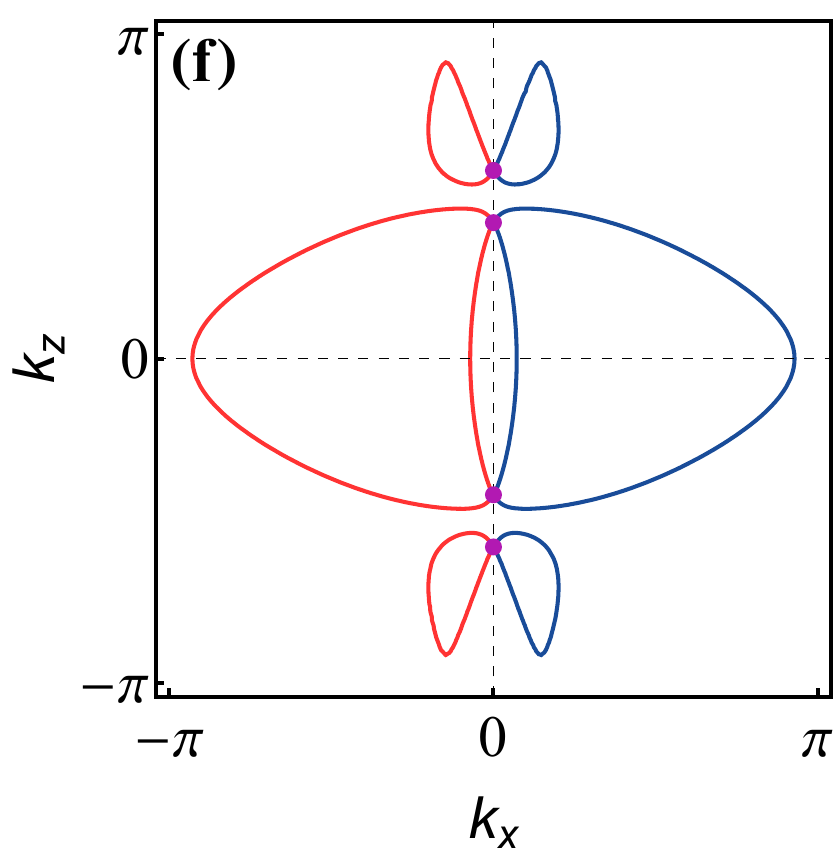}\label{fig:Bandf}}
  \caption{The quasiparticle spectrum and Bogoliubov Fermi surfaces of the WSC model. (a) The spectrum in the $k_y=0$ plane with $\Delta_0=0.3$ and $\boldsymbol{\eta}=0$, in which there are four BW nodes and two Bogoliubov nodal lines. (b) The spectrum along the $k_z$ axis with $\Delta_0=0.3$ and $\eta_x=0$, the green and red dashed lines correspond to $\eta_z=0.6$ and $\eta_z=1.2$, respectively, the gray line with $\boldsymbol{\eta}=0$. Along the line $(k_x,0,\arccos{(\Delta_0/\sqrt{1-\eta^2_z})})$, the spectra with $\boldsymbol{\eta}=(1.2,0,0)$ and $\boldsymbol{\eta}=(0.85,0,0.8)$ are plotted, by red dashed lines, in (c) and (d), respectively, where $\Delta_0=0.15$. The insets show the spectra along the line $(k_x,0,\arccos{(-\Delta_0 / \sqrt{1-\eta^2_z})})$ under the same parameters. Gray lines in (c) and (d) correspond to $\boldsymbol{\eta}=0$ and $\boldsymbol{\eta}=(0,0,0.8)$, respectively. (e) and (f) show the Bogoliubov Fermi surfaces corresponding to the red dashed lines in (c) and (d), respectively. The purple dots denote the type-II BW nodes at which the electron-like (red) and hole-like (blue) pockets touch.} \label{fig:BandS}
\end{figure}

When $\Delta_0=0$, the system is in the normal state, and the Hamiltonian~(\ref{EQ:HamBDG}) describes a WSM expressed in the electron-hole representation. In this representation, each Weyl node is doubled into an electron and hole node. Then, the pair potential $\Delta$ mixes an electron and hole node that originate from different Weyl nodes into two BW nodes lying in the superconducting gap, with their spacing increasing as $|\Delta_0|$ increases.  We note that this differs from the Meng-Balents model, in which the pair potential mixes an electron and hole node that originate from the same Weyl node, and each of the two pairs of BW cones is located at the same momentum point~\cite{Meng.PRB2012,Meng.PRB2017,Pacholski.PRL2018}. The Hamiltonian~(\ref{EQ:HamBDG}) also describes, besides the two pairs of BW nodes located at $(0,0,\pm \arccos{(\pm \Delta_0)})$ (in this work, we set $|\Delta_0|<1$), respectively, two nodal spherical surfaces with a radius of $k_r=\tfrac{\pi}{2}$ (an approximate spherical surface due to the underlying superlattice) and a center at ($0,0,0$), as we can see in Fig.~\ref{fig:Banda}. These two nodal surfaces coincide in momentum space but have opposite energies (they do not cross the Fermi level for $\boldsymbol{\eta}=0$) due to the particle-hole symmetry of Hamiltonian~(\ref{EQ:HamBDG}).

At low energies, the Weyl nodes in Hamiltonian~(\ref{EQ:HamWM}) can be described by a $2\times2$ linear Hamiltonian $H_W({\bf k}) = \pm v_F\hbar {\bf k}\cdot \boldsymbol{\sigma}$ (for simplicity, we have set an isotropic Weyl cone with Fermi velocity $v_x=v_y=v_z\equiv v_F$, which can always be achieved by rescaling the axes, say, $y^\prime=(v_x/v_y)y$). However, the Hamiltonian for a generic Weyl node should be written as~\cite{BurkovNatureMat2016}
\begin{eqnarray}
\mathcal{H}_W({\bf k}) = \hbar \boldsymbol{\eta} \cdot {\bf k} \sigma_0 \pm \hbar v_F {\bf k}\cdot \boldsymbol{\sigma},
\label{EQ:HamWNodal}
\end{eqnarray}
where the momentum ${\bf k}$ is measured from the Weyl point, $\boldsymbol{\eta}=(\eta_x,\eta_y,\eta_z)$ describes the tilt of the Weyl cone along the $k_x$, $k_y$ and $k_z$ directions, respectively. When $|\boldsymbol{\eta}|>v_F$, the two bands that touch at the Weyl node cross the Fermi level, forming electron and hole pockets, respectively. The Weyl node then becomes a point that connects an electron and a hole pocket, and such points are called type-II Weyl nodes~\cite{SoluyanovNature2015}.

Following Eq.~(\ref{EQ:HamWNodal}), we introduce a tilt term into the lattice Hamiltonian~(\ref{EQ:HamWM}), which is written as
\begin{eqnarray}
\Lambda({\bf k}) = t_0 (\eta_x \sin{k_x a_0} + \eta_z \cos{k_z a_0}) \sigma_0.
\end{eqnarray}
Without loss of generality, we have set $\eta_y=0$ in this work, as we can always orient the $x$-$y$ axes so that the Weyl cones are tilted in the $x$-$z$ plane. Therefore, the BdG Hamiltonian for the generic WSC is written as
\begin{eqnarray}
\mathcal{H}({\bf k})=
\begin{pmatrix}
\mathcal{H}_0({\bf k}) & \Delta \\
\Delta^* & -\sigma_y\mathcal{H}^*_0(-{\bf k})\sigma_y
\end{pmatrix},
\label{EQ:TiltBDG}
\end{eqnarray}
with $\mathcal{H}_0({\bf k}) = H_0({\bf k}) + \Lambda({\bf k})$, where $\Lambda({\bf k})$ can tilt the BW cones along the $k_x$ and $k_z$ directions.

Consider a nonzero tilt $\boldsymbol{\eta}$. The Hamiltonian~(\ref{EQ:TiltBDG}) describes four tilted BW cones located at ($0$, $0$, $\pm\arccos{(\pm\Delta_0/\sqrt{1-\eta^2_z})}$), respectively. Obviously, when $|\eta_z|>1$, the four BW nodes disappear, and the spectrum of Hamiltonian~(\ref{EQ:TiltBDG}) opens a gap, as shown in Fig.~\ref{fig:Bandb} by the red dashed lines. However, such a scenario will not occur for $|\eta_x|>1$. A nonzero $\eta_x$ tilts the dispersion of BW cones and nodal spherical surfaces along the $k_x$ direction. In contrast, a nonzero $\eta_z$ tilts the dispersion of BW cones along the $k_z$ direction while it lifts the degeneracy of nodal spherical surfaces, except for the two points $(0,0,\pm \tfrac{\pi}{2})$, as seen in Figs.~\ref{fig:Bandb},~\ref{fig:Bandc} and~\ref{fig:Bandd}. When $|\boldsymbol{\eta}|>1$ ($|\eta_z|<1$), the four nodes evolve into the type-II BW nodes, at each of which an electron-like and a hole-like pocket touch, and the superconductor shows Bogoliubov Fermi surfaces (In the WSC model, Bogoliubov Fermi surfaces emerge when $|\eta_x|+|\eta_z|>1$, while BW cones retain type-I when $|\boldsymbol{\eta}|<1$), as shown in Figs.~\ref{fig:Bande} and~\ref{fig:Bandf}. We also note that the inversion-symmetric WSC may feature topologically protected emergent Fermi surfaces in the presence of an Abrikosov vortex lattice~\cite{SongyangPu.Arxiv2024}.

\section{Landau levels in the mixed state}\label{sec:LandauL}
We assume the heterostructure WSC is the second-kind superconductor that hosts overlapping vortices at the intermediate magnetic field, $H_{c1}\ll B_0\ll H_{c2}$, where the magnetic field ${\bf B}=B_0\hat{z}$ is shown in Fig.~\ref{fig:hstrut}, with $H_{c1}$ and $H_{c2}$ being the lower and upper critical field, respectively. In this regime, vortex cores comprise a negligible fraction of the sample. Therefore, the magnitude of the order parameter is approximately uniform throughout the sample and equals $\Delta_0$, while the phase $\phi$ is strongly position-dependent. Here, we calculate the LLs of the generic WSC in a magnetic field ${\bf B}$ using two methods: one is an analytical approach employing the low-energy continuum model, and the other is a numerical approach studying the full tight-binding Hamiltonian on the vortex lattice. 

\subsection{Continuum formulation}\label{subsec:Continuum}
The perpendicular magnetic field ${\bf B}$ is produced by a vector potential ${\bf A}(x,y)$. The BdG Hamiltonian of the heterostructure WSC in this field is written as
\begin{eqnarray}
\mathcal{H}({\bf k})=
\begin{pmatrix}
\mathcal{H}_0({\bf k}-e{\bf A}) & \Delta_0e^{i\phi} \\
\Delta_0e^{-i\phi} & -\sigma_y\mathcal{H}_0^*(-{\bf k}-e{\bf A})\sigma_y
\end{pmatrix},
\label{EQ:HamMix}
\end{eqnarray}
where the single-particle Hamiltonian $\mathcal{H}_0$ is given by
\begin{eqnarray}
\begin{aligned}
\mathcal{H}_0({\bf k}) &= \hbar v_F \left[k_x\sigma_x + k_y\sigma_y + (K^2-k^2_z)\sigma_z\right] - \mu \sigma_0 + \Lambda({\bf k}), \\
\Lambda({\bf k}) &= \hbar \left[\eta_x k_x + \eta_z (K^2-k^2_z)\right] \sigma_0.
\end{aligned}
\end{eqnarray}
At zero magnetic field, Eq.~(\ref{EQ:HamMix}) shows four tilted BW nodes at $(0,0,\pm \left(K^2 \pm \Delta_0/ (1-\eta_z^2)^{1/2}\right)^{1/2})$, respectively. Following the method in Ref.~\onlinecite{Pacholski.PRL2018}, the phase factors $e^{i\phi}$ in the pair potential can be removed from the off-diagonal components and incorporated into the single-particle Hamiltonian $\mathcal{H}_0$. This is accomplished by a gauge transformation (Anderson gauge) $\mathcal{H} \to \mathcal{U}^\dagger \mathcal{H} \mathcal{U}$ with
\begin{eqnarray}
\mathcal{U} =
\begin{pmatrix}
e^{i\phi} & 0 \\
0 & 1
\end{pmatrix}.
\end{eqnarray}
 Setting $\hbar \equiv 1$, the transformed BdG Hamiltonian
\begin{eqnarray}
\tilde{\mathcal{H}} ({\bf k})=
\begin{pmatrix}
\mathcal{H}_0 ({\bf k}+{\bf a}+m{\bf v}_s) & \Delta_0 \\
\Delta_0 & -\sigma_y \mathcal{H}_0^*(-{\bf k}-{\bf a}+m{\bf v}_s)\sigma_y
\end{pmatrix},
\label{EQ:GHamMix}
\end{eqnarray}
with the definitions
\begin{eqnarray}
{\bf a}=\tfrac{1}{2}\nabla\phi, \qquad m{\bf v}_s=\tfrac{1}{2}\nabla\phi-e{\bf A}.
\end{eqnarray}
Both the gauge field ${\bf a}(x,y)$ and the supercurrent velocity ${\bf v}_s(x,y)$ wind around the vortex cores, at positions ${\bf R}_n$, due to
\begin{eqnarray}
\nabla\times\nabla\phi=2\pi\hat{z}\sum_{n}\delta({\bf r}-{\bf R}_n).
\label{EQ:PhaseW}
\end{eqnarray}
Besides, the supercurrent velocity should have vanishing divergence, i.e., $\nabla\cdot\nabla\phi=0$ (we choose $\nabla\cdot{\bf A}=0$).

Rotating the spin, we can transform Hamiltonian~(\ref{EQ:GHamMix}) into block diagonal form by $\tilde{\mathcal{H}} \to \mathcal{V}^{\dagger} \tilde{\mathcal{H}} \mathcal{V}$, with
\begin{eqnarray}
\mathcal{V}=\exp{(\tfrac{1}{2}i\vartheta \nu_y \sigma_z)}, \quad \tan \vartheta = -\tfrac{\Delta_0}{v_F k_z}, \quad \vartheta \in (0, \pi),
\end{eqnarray}
where $\boldsymbol{\nu}$ is the Pauli matrix that acts on the electron-hole index. At low energies, the rotation gives rise to an effective $2\times 2$ Hamiltonian, which is written as
\begin{eqnarray}
\begin{aligned}
\mathcal{H}_{\pm}({\bf k}) &= v_F\sum_{\alpha=x,y} [\pm (k_\alpha + a_\alpha) + \kappa mv_{s,\alpha}]\sigma_\alpha \\
&+ (K^2-k_z^2\mp \tfrac{\Delta^2}{\Gamma})\sigma_z \mp \kappa \mu \sigma_0 + \Lambda_\pm ({\bf k}), \\
\Lambda_\pm ({\bf k}) &= [\eta_x(k_x+a_x\pm \kappa v_{s,x}) \pm \kappa \eta_z(K^2-k_z^2)]\sigma_0,
\end{aligned}
\label{EQ:LNHam}
\end{eqnarray}
where $\Gamma=\sqrt{\Delta^2+v_F^2k_z^2}$ and $\kappa=-v_Fk_z/\Gamma$. Usually, the superconducting pair potential satisfies $\Delta_0 \ll t_0$; if we consider a strong magnetic field, the terms containing $\Delta_0$ in the off-diagonal block of the rotated Hamiltonian become irrelevant, and thus does not mix the LLs of the electron and hole cones.

Considering  $\mu=0$, let's take $\mathcal{H}_{+}({\bf k})$ as for example. Around a BW node, we can expand $\mathcal{H}_{+}({\bf k})$ to the linear term of $k_z$, resulting in
\begin{eqnarray}
\mathcal{H}_{BW} = v_F \boldsymbol{\Pi} \cdot \boldsymbol{\sigma} + \kappa \eta_z \bar{k}_z \sigma_0 + \eta_x (k_x + e\mathcal{A}_x) \sigma_0,
\label{EQ:BWBHam}
\end{eqnarray}
where $\boldsymbol{\Pi}={\bf k}+e\boldsymbol{\mathcal{A}}$ with $e\boldsymbol{\mathcal{A}}={\bf a} +\kappa m\boldsymbol{v_s}$ ($\mathcal{A}_z=0$). Here, $\bar{k}_z$ is measured from the BW point. We dropped the coefficients of $\bar{k}_z$, which can be compensated for by scaling the $\bar{k}_z$ axis. Furthermore, when expanding $K^2-k_z^2 - \tfrac{\Delta_0^2}{\Gamma}$ and $K^2-k_z^2$ at the BW node, the resulting coefficients of $k_z$ have a slight difference that can be neglected because $\Delta^2_0 \ll \Gamma$. For the case of $\boldsymbol{\eta}=0$, Pacholski \textit{et al.} have proved that the Hamiltonian~(\ref{EQ:BWBHam}) can exhibit Dirac-LLs in the effective magnetic field $\mathcal{B}=\nabla \times \mathcal{A}$~\cite{Pacholski.PRL2018}.

With a nonzero tilt $\boldsymbol{\eta}$, the term $\eta_x e\mathcal{A}_x \sigma_0$ can be equivalent to the effect of an in-plane electric field; for instance, $E_{eff}=\eta_x \mathcal{B}$ in the negative $y$ direction when using the Landau gauge $\mathcal{A}=(-\mathcal{B}y,0,0)$. The eigenvalues of Hamiltonian~(\ref{EQ:BWBHam}) can be solved through a lengthy algebraic calculation~\cite{PeresJPCM2007}. Alternatively, we can rewrite the eigenvalue equation $\mathcal{H}_{BW}\Psi=E\Psi$ as
\begin{eqnarray}
e^{\tfrac{\theta}{2}\sigma_x} \mathcal{H}_{BW} e^{\tfrac{\theta}{2}\sigma_x} \tilde{\Psi} = E e^{\theta\sigma_x} \tilde{\Psi}
\label{EQ:HeigenEQ}
\end{eqnarray}
using a hyperbolic transformation~\cite{TchoumakovPRL2016}, where $\tilde{\Psi}=\mathcal{N}e^{-\tfrac{\theta}{2}\sigma_x} \Psi$ with $\mathcal{N}$ being a normalization constant. The term containing $\Pi_x$ in the diagonal elements of the transformed Hamiltonian becomes zero when we set $\tanh{\theta} = \zeta = -\eta_x/v_F$, which in turn eliminates the tilted term with $\eta_x$ (the in-plane electric field). Then the equation~(\ref{EQ:HeigenEQ}) can be reduced to
\begin{eqnarray}
v_F(\tilde{\Pi}_x \sigma_x + \Pi_y \sigma_y + \bar{k}_z \sigma_z) \tilde{\Psi} = \gamma(E-\kappa\eta_z \bar{k}_z)\tilde{\Psi},
\label{EQ:SimHeigEQ}
\end{eqnarray}
where $\gamma=(1-\zeta^2)^{-1/2}$, $\tilde{\Pi}_x=\Pi_x/\gamma + \gamma\zeta(\kappa \eta_z \bar{k}_z-E)/v_F$, with a reduced effect magnetic field $\tilde{\mathcal{B}}=\mathcal{B}/\gamma$ due to $[\tilde{\Pi}_x,\Pi_y]=-i/l^2_{\tilde{\mathcal{B}}}=-i/\gamma l^2_{\mathcal{B}}$. This is equivalent to a Lorentz boost in the $x$ direction with the relativistic parameter $\tanh{\theta} = \zeta$~\cite{LukosePRL2007,YuZhiMingPRL2016}.

When $|\eta_x/v_F|<1$, the system is in the magnetic regime due to the vanishing in-plane electric field, in which the cyclotron orbits are still closed. Thus, we can easily obtain the LLs of Hamiltonian~(\ref{EQ:BWBHam}) by Eq.~(\ref{EQ:SimHeigEQ}), reading
\begin{eqnarray}
\begin{aligned}
E_{n,\pm}(\bar{k}_z) &= \kappa \eta_z \bar{k}_z \pm \tfrac{1}{\gamma} \sqrt{v_F^2 \bar{k}^2_z + 2v_F^2 e\mathcal{B}n/\gamma}, \quad n \ge 1, \\
E_0(\bar{k}_z) &= \kappa \eta_z \bar{k}_z - \tfrac{1}{\gamma} v_F \bar{k}_z, \quad n=0,
\end{aligned}
\label{EQ:LLs}
\end{eqnarray}
with reduced LL spacings due to $\tilde{\mathcal{B}}=\mathcal{B}\sqrt{1-(\eta_x/v_F)^2}$. When $|\eta_x/v_F|>1$, it is obvious that $\gamma$ is imaginary, resulting in the LLs collapse. Consequently, the system transitions into the electric regime, characterized by open orbits that prevent LL quantization.

\subsection{Lattice formulation}\label{subsec:Lattice}
In the mixed state, the LL spectrum of the WSC can be obtained by tight-binding calculations on a vortex lattice. In the $x$-$y$ plane, the vortex lattice possesses a 2D square magnetic unit cell $l_B\times l_B$, and each magnetic unit cell includes two singly quantized vortices, each of which carrying flux $hc/2e$, as shown in Fig.~\ref{fig:Lattice}. The magnetic length $l_B=Na_0$ is defined as $l_B=\sqrt{\phi_0/B_0}$, with the flux quantum $\phi_0=hc/e$. For simplicity, we set $c=e=1$ and $N=4n+2$, where $n$ is a positive integer, in our numerical calculations~\cite{Vafek.PRL2006,Vafek.PRB63.2001}. The superconducting phase $\phi({\bf r})\equiv \phi({x,y})$ on the vortex lattice can be found in Ref.~\onlinecite{Melikyan.PRB2007}, where it is expressed in closed form through the Weierstrass sigma function. Around each vortex, the superconducting phase $\phi({\bf r})$ undergoes a $2\pi$ winding, as shown in Eq.~(\ref{EQ:PhaseW}) in the continuum description.

In the magnetic field ${\bf B}$, the tight-binding Hamiltonian of the heterostructure WSC is written as
\begin{eqnarray}
\begin{aligned}
\mathcal{H}_{TB} = &\sum_{\langle{\bf r}{\bf r^\prime}\rangle \sigma \sigma^\prime} \left( - \tilde{t}^{\sigma \sigma^\prime}_{{\bf r}{\bf r^\prime}} c^{\dagger}_{{\bf r} \sigma} c_{{\bf r^\prime} \sigma^\prime} + \mathrm{H.c.}\right) + (\lambda \mathfrak{B}^\sigma -\mu) \sum_{{\bf r} \sigma} c^{\dagger}_{{\bf r} \sigma} c_{{\bf r}  \sigma} \\
& + \sum_{\bf r} \left(\Delta_{\bf r} c^{\dagger}_{{\bf r} \uparrow} c^{\dagger}_{{\bf r} \downarrow} + \mathrm{H.c.}\right),
\end{aligned}
\label{EQ:TBHam}
\end{eqnarray}
where the sum is over the nearest neighbors $\langle{\bf r}{\bf r^\prime}\rangle$ of the tight-binding lattice. Here, $\sigma$ denotes the spin, $\mathfrak{B}^\uparrow=\beta$ and $\mathfrak{B}^\downarrow=-\beta$, $\mu$ is the chemical potential. The hopping integral $\tilde{t}^{\sigma \sigma^\prime}_{{\bf r}{\bf r^\prime}} = t^{\sigma \sigma^\prime}_{{\bf r}{\bf r^\prime}} \exp{(-i{\bf A}_{{\bf r}{\bf r^\prime}})}$. Considering the symmetric gauge, the magnetic vector potential ${\bf A}_{{\bf r}{\bf r^\prime}}$ in the Peierls factor can be written as ${\bf A}_{{\bf r}{\bf r}+\hat{x}}=-\pi y\Phi/\phi_0$ and ${\bf A}_{{\bf r}{\bf r}+\hat{y}}=\pi x\Phi/\phi_0$, with $\Phi$ being the magnetic flux through an elementary plaquette. In the heterostructure superconductor, the pairing field is given by $\Delta_{\bf r} = \Delta_0e^{i\phi({\bf r})}$, where the ansatz $e^{i\phi({\bf r})}$ for the $s$-wave pair phase factors is obtained through a self-consistent calculation.

The Hamiltonian~(\ref{EQ:TBHam}) can be diagonalized by solving the BdG equation $\hat{\mathcal{H}}\psi_{\bf r}=E\psi_{\bf r}$, where the lattice operator is written as
\begin{eqnarray}
\hat{\mathcal{H}} =
\begin{pmatrix}
\hat{\boldsymbol{\varepsilon}}_{\bf r} + h_0 & -i \sigma_y \hat{\Delta}_{\bf r} \\
i \sigma_y \hat{\Delta}_{\bf r}^* & - \hat{\boldsymbol{\varepsilon}}^*_{\bf r} - h_0
\end{pmatrix}
\label{EQ:Latop}
\end{eqnarray}
with $h_0 = \beta \sigma_z - \mu \sigma_0$, $\hat{\boldsymbol{\varepsilon}}_{\bf r}$ is a $2\times 2$ matrix with elements $(\hat{\boldsymbol{\varepsilon}}_{\bf r})_{\sigma \sigma^\prime}$. The BdG Hamiltonian acts on the Nambu spinor $\psi_{\bf r}$, and for a wave function at the lattice site ${\bf r}$, the operators $(\hat{\boldsymbol{\varepsilon}}_{\bf r})_{\sigma \sigma^\prime}$ and $\hat{\Delta}_{\bf r}$ are defined respectively as
\begin{subequations}
 \begin{align}
 (\hat{\boldsymbol{\varepsilon}}_{\bf r})_{\sigma \sigma^\prime} f_{\bf r} & = \sum_{\boldsymbol{\delta}} (\mathcal{T}_{\boldsymbol{\delta}})_{\sigma \sigma^\prime} \times e^{-i{\bf A}_{{\bf r}{\bf r} + \boldsymbol{\delta}}} f_{{\bf r} + \boldsymbol{\delta}}, \label{Ophop}\\
 \hat{\Delta}_{\bf r} f_{\bf r} &= \Delta_0e^{i\phi({\bf r})} f_{\bf r},
 \end{align}
 \label{Operators}%
\end{subequations}
where the unit vector $\boldsymbol{\delta} = \pm \hat{\boldsymbol{x}}, \pm \hat{\boldsymbol{y}}, \pm \hat{\boldsymbol{z}}$ points to the six nearest neighbors in the lattice. In the heterostructure model, the hopping matrix is given by $\mathcal{T}_{\boldsymbol{\delta}} = - \tfrac{t_0}{2} (\varkappa_{\boldsymbol{\delta}} \boldsymbol{\sigma} \cdot \boldsymbol{\delta} + {\dot \varkappa}_{\boldsymbol{\delta}} {\sigma}_z + {\ddot \varkappa}_{\boldsymbol{\delta}} {\sigma}_0 )$, where $\varkappa_{\pm x,\pm y} = i$, $\varkappa_{\pm z} = 0$,  ${\dot \varkappa}_{\pm x,\pm y} = -\lambda$, ${\dot \varkappa}_{\pm z} = -1$, ${\ddot \varkappa}_{\boldsymbol{\delta}} = i \boldsymbol{\eta} \cdot \boldsymbol{\delta}$ for $\boldsymbol{\delta}=\pm x,\pm y$ and ${\ddot \varkappa}_{\boldsymbol{\delta}} =1$ for $\boldsymbol{\delta}=\pm z$. Although the vortices are periodically configured in the vortex lattice, the Hamiltonian~(\ref{EQ:Latop}) remains invariant only when the discrete translations are accompanied by a gauge transformation (magnetic translations).

As shown in Refs.~\cite{Franz.PRL2000,Vafek.PRL2006,Melikyan.PRB2007,Vafek.PRB63.2001,WangLuyangPRB2013}, the gauge transformation matrix (symmetric gauge) that transforms $\hat{\mathcal{H}}$ into a periodic Hamiltonian $\tilde{\mathcal{H}} = \mathcal{U}^\dagger \hat{\mathcal{H}} \mathcal{U}$ is given by
\begin{eqnarray}
\mathcal{U}=
\begin{pmatrix}
e^{\frac{i}{2}\phi({\bf r})} & 0 \\
0 & e^{- \frac{i}{2}\phi({\bf r})}
\end{pmatrix}.
\label{EQ:SGTM}
\end{eqnarray}
It should be noted that, on the vortex lattice, the symmetric transformation~(\ref{EQ:SGTM}) with the phase factors $e^{\pm\frac{i}{2}\phi({\bf r})}$ is not single-valued due to the $2\pi$ winding of the superconducting phase $\phi({\bf r})$ around each vortex. Therefore, the resulting Hamiltonian is also not single-valued. Such multiple valuedness could be handled by introducing compensating branch cuts, as shown in Fig.~\ref{fig:Lattice}, in which each branch cut connects to two vortices in a magnetic cell, thus preserving the periodicity of the system.

Specifically, the multiple valuedness of the phase factors in the transformed Hamiltonian $\tilde{\mathcal{H}}$ could be eliminated by using the equation
\begin{eqnarray}
e^{- \frac{i}{2}\phi({\bf r})} e^{\frac{i}{2}\phi({\bf r^\prime})} e^{-i{\bf A}_{{\bf r}{\bf r^\prime}}} = z_{2,{\bf r}{\bf r^\prime}} \times e^{iV_{{\bf r}{\bf r^\prime}}},
\end{eqnarray}
where
\begin{subequations}
 \begin{align}
z_{2,{\bf r}{\bf r^\prime}} &= \tfrac{e^{i\phi({\bf r})} + e^{i\phi({\bf r^\prime})}}{|e^{i\phi({\bf r})} + e^{i\phi({\bf r^\prime})}|} \times e^{- \frac{i}{2}\phi({\bf r})} e^{-\frac{i}{2}\phi({\bf r^\prime})}, \\
e^{iV_{{\bf r}{\bf r^\prime}}} &= \tfrac{1+e^{i[\phi({\bf r^\prime}) - \phi({\bf r})]}}{|1+e^{i[\phi({\bf r^\prime}) - \phi({\bf r})]}|} e^{-i{\bf A}_{{\bf r}{\bf r^\prime}}}.
\end{align}
\end{subequations}
Obviously,  the factor $e^{iV_{{\bf r}{\bf r^\prime}}}$ is single-valued. The multiple valuedness of the transformed Hamiltonian is transferred into the $Z_2$ field that only has values of $1$ and $-1$, depending on $\phi({\bf r})$ and $\phi({\bf r^\prime})$. Therefore we will set $z_{2,{\bf r} {\bf r^\prime}}=1$ on each bond except the ones crossing the branch cut where $z_{2,{\bf r} {\bf r^\prime}}=-1$.

The transformed Hamiltonian is now
\begin{eqnarray}
\tilde{\mathcal{H}} =
\begin{pmatrix}
\tilde{\boldsymbol{\varepsilon}}_{\bf r} + h_0 & -i \sigma_y \tilde{\Delta}_{\bf r} \\
i \sigma_y \tilde{\Delta}_{\bf r}^* & - \tilde{\boldsymbol{\varepsilon}}^*_{\bf r} - h_0
\end{pmatrix},
\end{eqnarray}
where the transformed lattice operators satisfy
\begin{subequations}
 \begin{align}
 (\tilde{\boldsymbol{\varepsilon}}_{\bf r})_{\sigma \sigma^\prime} u_{\bf r} &= \sum_{\boldsymbol{\delta}} (\mathcal{T}_{\boldsymbol{\delta}})_{\sigma \sigma^\prime} \times z_{2,{\bf r}{\bf r}+\boldsymbol{\delta}} \times e^{i V_{{\bf r}{\bf r} + \boldsymbol{\delta}}} u_{{\bf r} + \boldsymbol{\delta}},\\
 \tilde{\Delta}_{\bf r} u_{\bf r} &= \Delta_0 u_{\bf r}.
 \end{align}
\end{subequations}
Due to the periodicity of $V_{{\bf r}{\bf r^\prime}}$~\cite{NielsenAPS1995} and the periodic arrangement of branch cuts on the vortex lattice, the resulting Hamiltonian $\tilde{\mathcal{H}}$ is invariant under discrete translations by the vortex lattice constant $l_B$. Thus, it can be diagonalized in the Bloch basis. By extracting the crystal wave vector ${\bf k}$ from the Bloch wave functions, we get $\mathscr{H}({\bf k})=e^{-i{\bf k}{\bf r}} \tilde{\mathcal{H}} e^{i{\bf k}{\bf r}}$ acting on the Hilbert space of periodic Nambu spinors.

\begin{figure}[ht]
  \centering
  \subfigure{\includegraphics[width=1.6in]{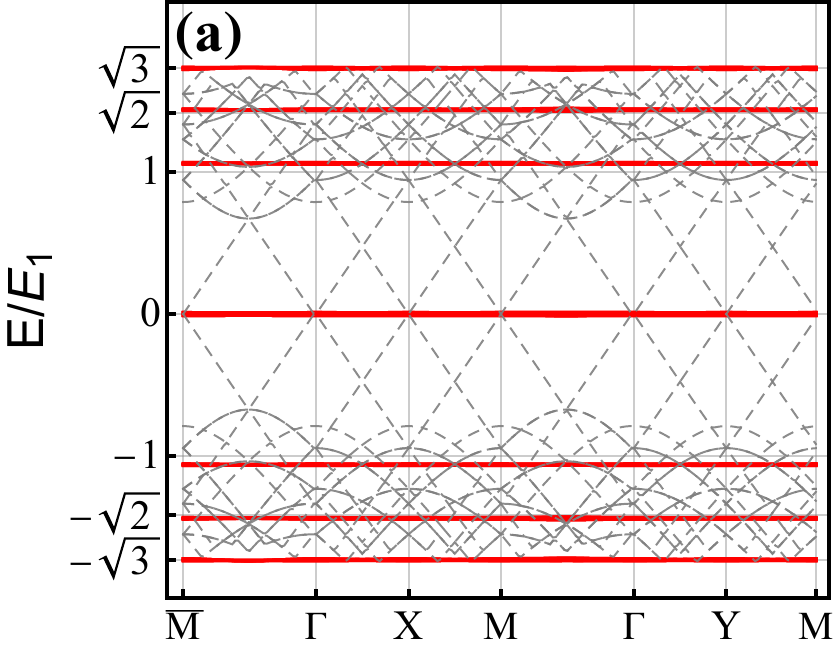}\label{fig:LLa}}~~
  \subfigure{\includegraphics[width=1.6in]{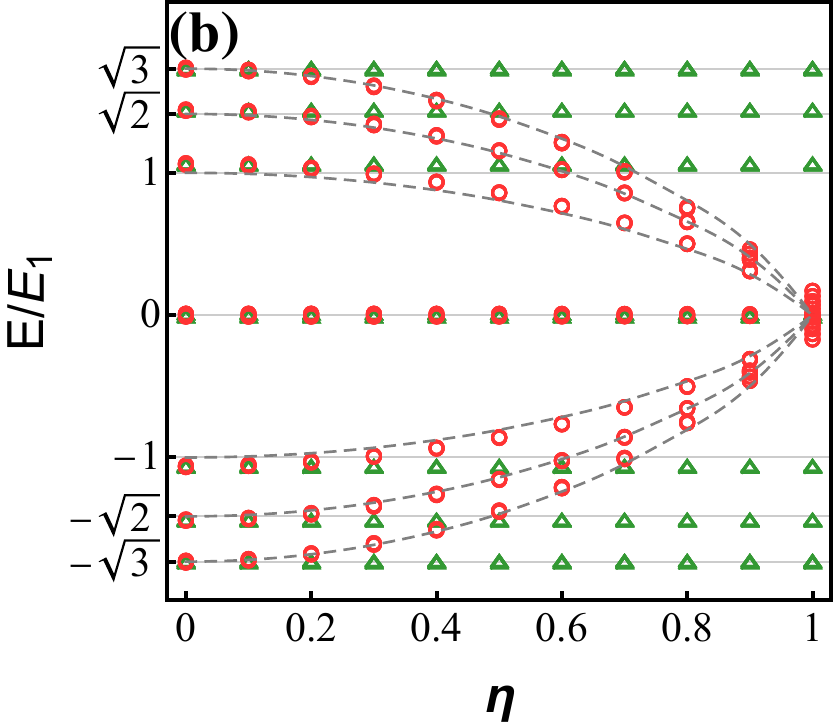}\label{fig:LLb}}
  \subfigure{\includegraphics[width=1.6in]{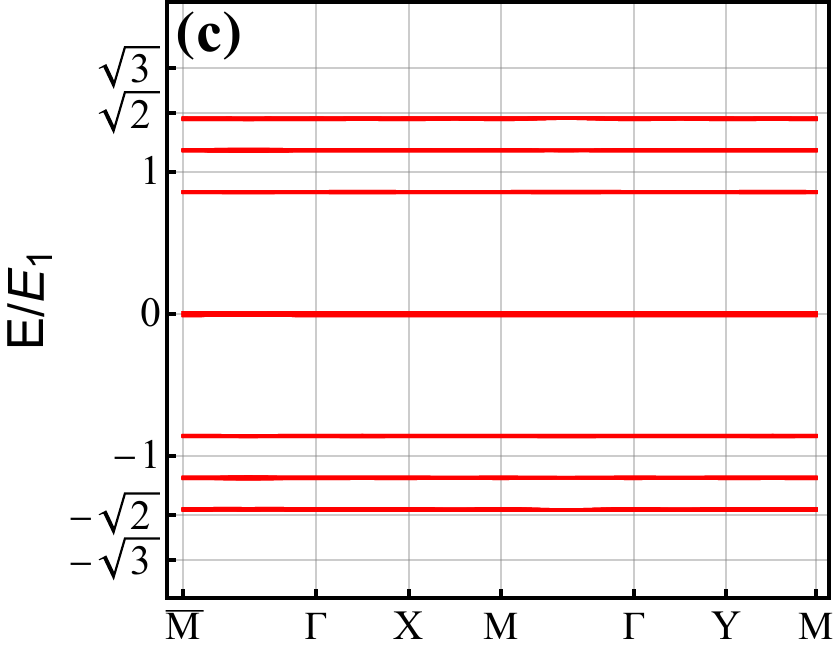}\label{fig:LLc}}~~
  \subfigure{\includegraphics[width=1.6in]{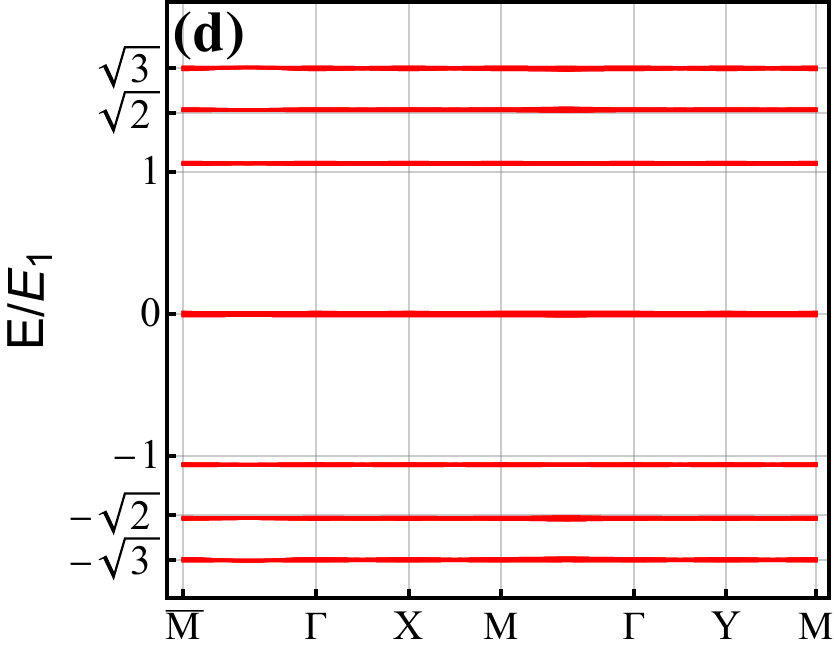}\label{fig:LLd}}
  \subfigure{\includegraphics[width=1.6in]{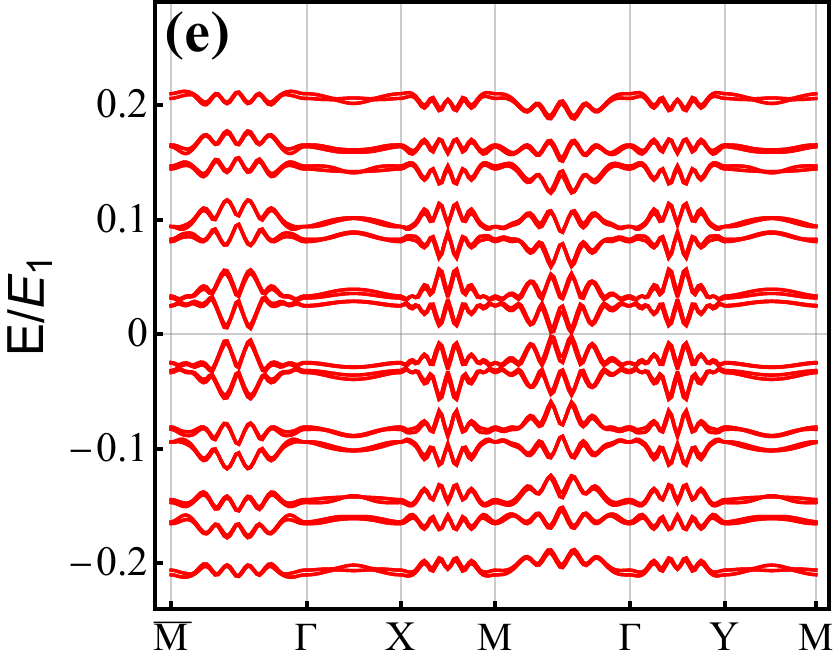}\label{fig:LLe}}~~
  \subfigure{\includegraphics[width=1.6in]{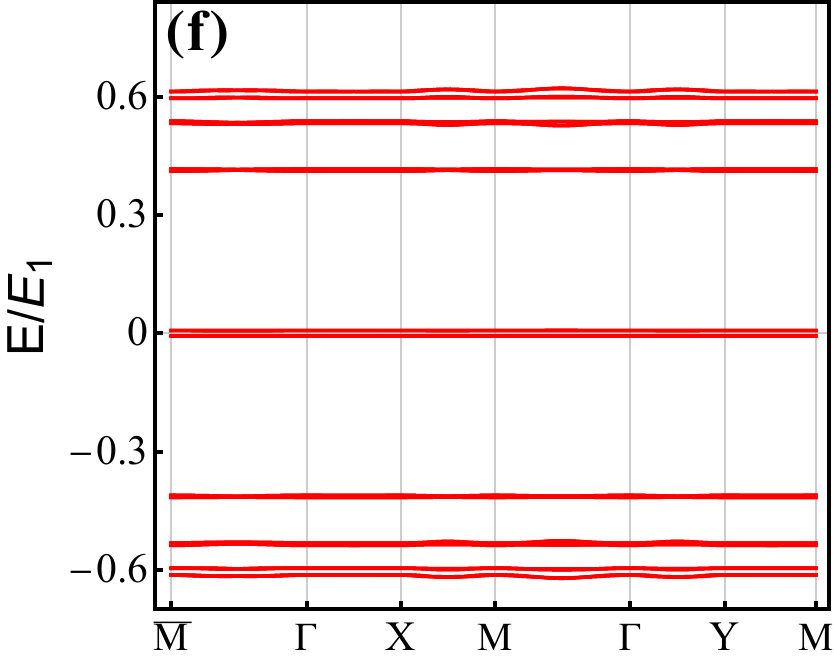}\label{fig:LLf}}
  \caption{The quasiparticle spectrum in the mixed state of the generic WSC with $\lambda=0$, $l_B=14$, $k_z=\tfrac{\pi}{2}$ and $\Delta_0=0.1$. (a) The LLs for the BW cone with $\boldsymbol{\eta}=0$, the black dashed curves (with $l_B=13$, $k_z=\arccos{(0.1)}$) denote the excitation spectrum of the WSC in zero magnetic field. (b) Energies for the first few LLs (with $k_x=k_y=0$) versus $\boldsymbol{\eta}$, the red circles and green triangles correspond to $\eta_z=0$ and $\eta_x=0$, respectively. The black dashed curves depict $E_n(\bar{k}_z=0)/E_1 = \sqrt{n/\gamma^3} = \sqrt{n}(1-(\eta_x/v_F)^2)^{3/4}$, where $E_1=2\sqrt{\pi}v_F/l_B$ with $v_F=1$. (c) and (d) show the LLs with $\boldsymbol{\eta}=(0.5,0,0)$ and $\boldsymbol{\eta}=(0,0,0.5)$, respectively, corresponding to the tilted type-I BW cones. (e) and (f) show the spectra with $\boldsymbol{\eta}=(1.2,0,0)$ and $\boldsymbol{\eta}=(0.85,0,0.8)$, respectively, corresponding to the type-II BW cones.} \label{fig:LLS}
\end{figure}
\begin{figure}[t]
  \centering
  \subfigure{\includegraphics[width=1.6in]{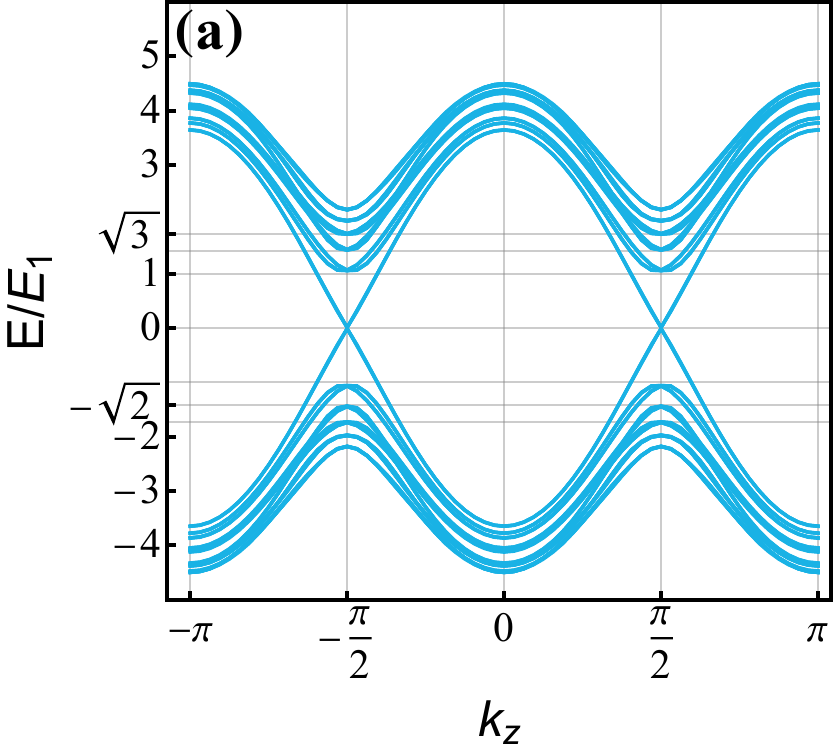}\label{fig:LLZa}}~~
  \subfigure{\includegraphics[width=1.6in]{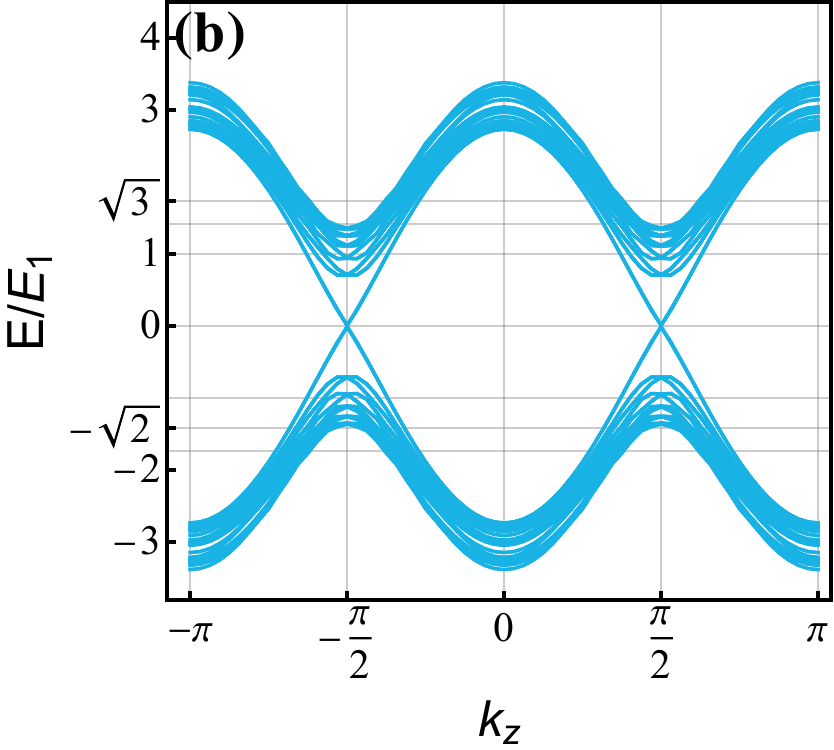}\label{fig:LLZb}}
  \subfigure{\includegraphics[width=1.6in]{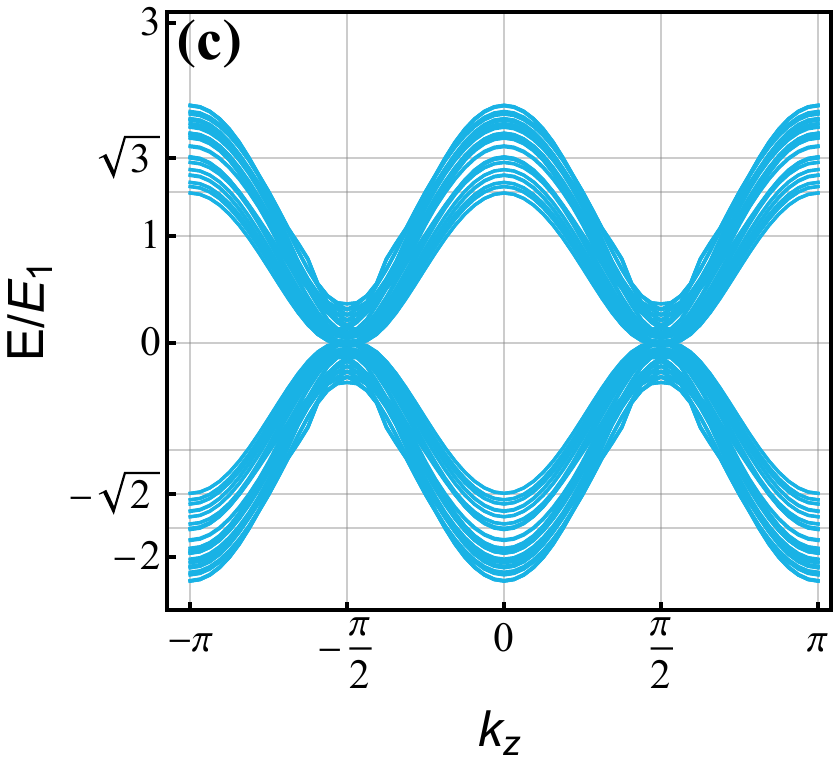}\label{fig:LLZc}}~~
  \subfigure{\includegraphics[width=1.6in]{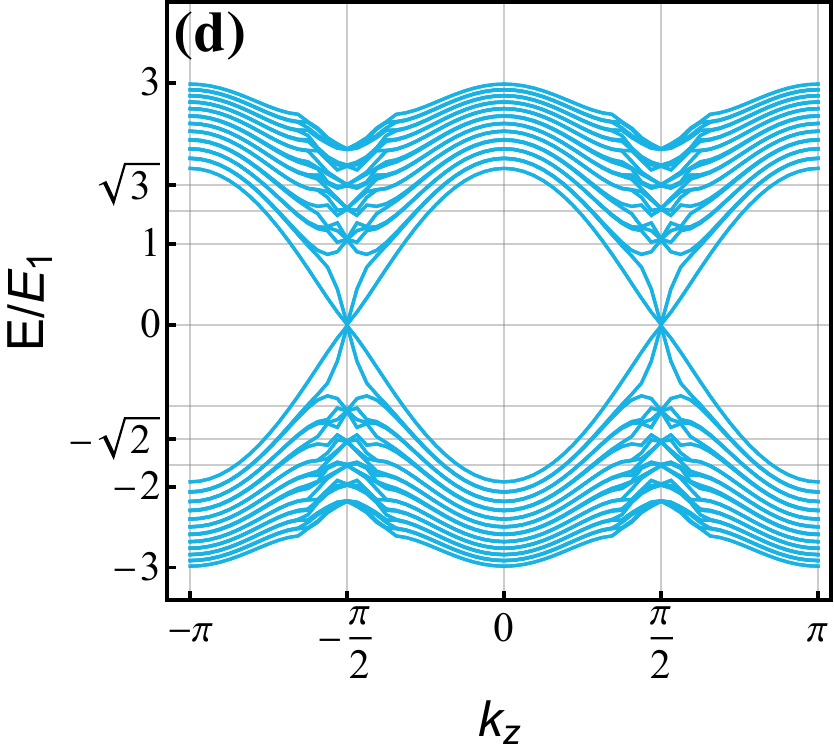}\label{fig:LLZd}}
  \subfigure{\includegraphics[width=1.6in]{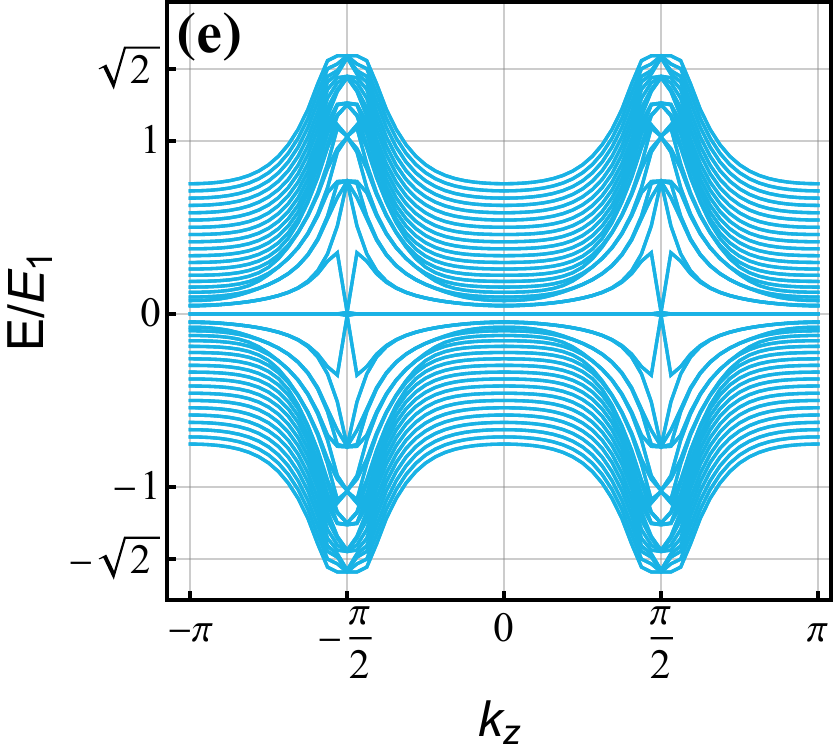}\label{fig:LLZe}}~~
  \subfigure{\includegraphics[width=1.6in]{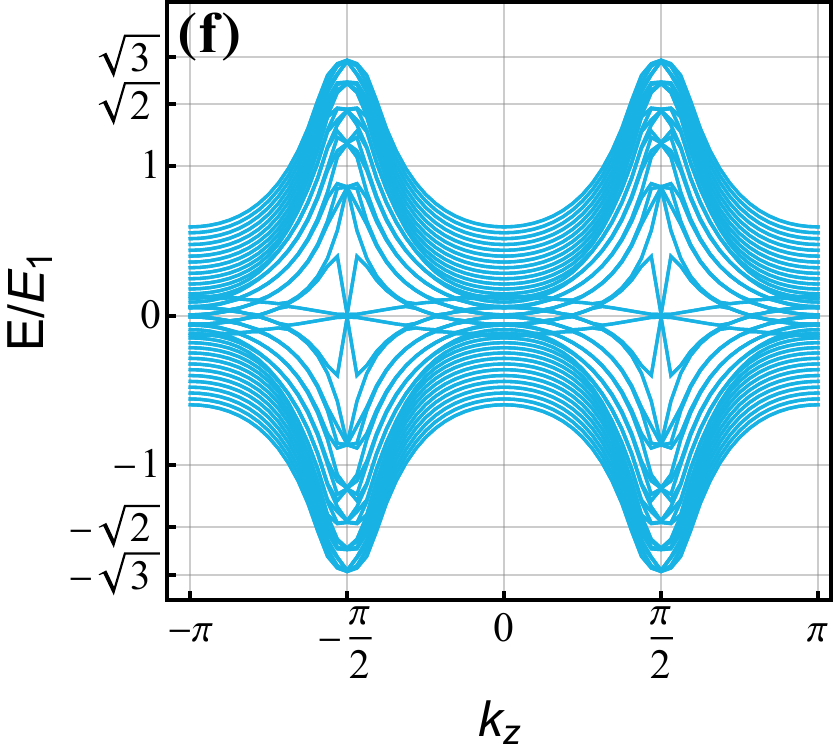}\label{fig:LLZf}}
  \caption{The quasiparticle spectrum along the momentum $k_z$ in the mixed state of the generic WSC with $\lambda=0$, $l_B=14$, $k_x=k_y=0$ and $\Delta_0=0.1$. (a) The spectrum for a BW cone with $\boldsymbol{\eta}=0$. (b), (c), (d), (e) and (f) show the spectra with $\boldsymbol{\eta}=(0.65,0,0)$, $\boldsymbol{\eta}=(1,0,0)$, $\boldsymbol{\eta}=(0,0,0.5)$, $\boldsymbol{\eta}=(0.6,0,0.8)$ and $\boldsymbol{\eta}=(0.5,0,0.9)$, respectively.} \label{fig:LLZS}
\end{figure}

The numerically calculated quasiparticle spectra in the mixed state of the generic WSC, using the full tight-binding Hamiltonian~(\ref{EQ:TBHam}), are shown in Figs.~\ref{fig:LLS} and~\ref{fig:LLZS}. For clarity, here we consider the $\lambda=0$ case. In this case, when $\Delta_0=0$ (WSM phase), there are four Weyl nodes in each of the two planar BZs $k_z=\pm \tfrac{\pi}{2}$, located at their center and corners, respectively. In the superconducting phase, as analyzed above, each Weyl node evolves into two BW nodes due to the pair potential. Therefore, in zero magnetic field, there are four BW nodes in each of the four planar BZs $k_z=\pm \arccos{(\pm \Delta_0)}$ for the WSC model with $\lambda=0$, as shown in Fig.~\ref{fig:LLa} by the black dashed lines. However, the tilt of these BW cones under $\boldsymbol{\eta}$ is the same as that in the $\lambda=1$ case, as seen the Fig.~\ref{fig:BandS}.

Fig.~\ref{fig:LLa} shows the spectrum in the mixed state of the WSC with untilted BW cones. At low energies, these LLs are consistent with the expectation $\sqrt{n}E_1$. Note that this LL spectrum is plotted in the $k_z=\tfrac{\pi}{2}$ plane (corresponding to $\arccos{(0)}$ rather than $\arccos{(\pm \Delta_0)}$), in which each level is eight-fold degenerate, coming from the eight BW cones in the $k_z=\arccos{(\pm \Delta_0)}$ planes. Fig.~\ref{fig:LLb} exhibits the energies of the first few LLs (with $k_x=k_y=0$ and $k_z=\tfrac{\pi}{2}$) versus $\eta_x$ and $\eta_z$ by the red circles and green triangles, respectively. The black dashed curves show the analytical results expressed in Eq.~(\ref{EQ:LLs}). We can see that $\eta_x$ squeezes the LL spacings while $\eta_z$ does not. Two concrete examples are shown in Figs.~\ref{fig:LLc} and~\ref{fig:LLd}, corresponding to the tilt parameters $\boldsymbol{\eta}=(0.5,0,0)$ and $\boldsymbol{\eta}=(0,0,0.5)$, respectively. The spectra in the mixed state of the WSC with type-II BW cones are shown in Figs.~\ref{fig:LLe} and~\ref{fig:LLf}, corresponding to the tilt parameters $\boldsymbol{\eta}=(1.2,0,0)$ and $\boldsymbol{\eta}=(0.85,0,0.8)$, respectively. When the projection $|\eta_{\bot}|\equiv|\eta_x|$ of the tilt in the plane perpendicular to the magnetic field is greater than $1$, the LLs collapse, as shown in Fig.~\ref{fig:LLe}; then the spectrum shows dispersing quasiparticle bands in all momentum directions. In contrast, the spectrum for the type-II BW cones with $|\eta_{\bot}|<1$, see Fig.~\ref{fig:LLf}, always exhibits LLs that are dispersionless in the $k_x$-$k_y$ plane, with decreased spacings due to the nonzero $\eta_x$.

Fig.~\ref{fig:LLZS} shows the LL spectrum as a function of the momentum $k_z$ for the WSC with a tilt parameter $\boldsymbol{\eta}$, where $k_x=k_y=0$. First, we plot the LLs for the case of $\boldsymbol{\eta}=0$ in Fig.~\ref{fig:LLZa}. Obviously, it is similar to that of the WSM, where each zeroth LL evolves into two quasiparticle chiral levels due to the superconducting pair. Figs.~\ref{fig:LLZb} and~\ref{fig:LLZc} show the LL spectrum of the WSC model with tilt parameters $\boldsymbol{\eta}=(0.65,0,0)$ and $\boldsymbol{\eta}=(1,0,0)$, respectively. We see once again that, as $\eta_x$ increase, the LL spacings are squeezed and collapse when $\eta_x=1$. The scenario is also reported in WSMs, see the \textit{Supplemental Material} of Ref.~\onlinecite{YuZhiMingPRL2016}. Figs.~\ref{fig:LLZd},~\ref{fig:LLZe} and~\ref{fig:LLZf} show the LL spectrum of the WSC with nonzero $\eta_z$, as elucidated by Eq.~(\ref{EQ:LLs}),  the LLs are tilted in the $k_z$ direction with a coefficient $\kappa\eta_z$, while the spacings retain invariant at $k_z=\tfrac{\pi}{2}$ when $\eta_x=0$, as shown in Fig.~\ref{fig:LLZd}. At the critical case with $|\boldsymbol{\eta}|=1$, as seen in Fig.~\ref{fig:LLZe}, the chiral zeroth LLs become flat in the spectrum. With a further increase in $\eta_z$, some LLs cross the Fermi level at certain $k_z$ points, corresponding to the type-II BW cones, as shown in Fig.~\ref{fig:LLZf}.

\section{Optical conductivity in the mixed state}\label{sec:MOconductivity}
LLs in the normal state can be typically detected through experiments such as quantum oscillations of various physical quantities and quantum Hall effect. However, to observe LLs in the superconducting state, most of these experiments can hardly work or may give elusive results, because of two reasons: (1) due to $U(1)$ symmetry breaking, charge is not conserved, making experiments relying on charge conservation to produce understandable results disadvantageous; and (2) due to particle-hole symmetry, LLs of Bogoliubov quasiparticles lie symmetrically about the Fermi level which is always zero, so that LLs cannot cross the Fermi level unless Bogoliubov Fermi surfaces exist~\cite{WangLuyangPRB2013}.

An alternative way to observe the LLs in the mixed state of superconductors is the optical conductivity measurement. Magneto-optical conductivity has been shown to be a powerful tool in the studies of topological semimetals. For instance, the longitudinal magneto-optical conductivity of graphene displays a series of delta-peaks that reflect the dispersionless LLs of the massless chiral fermions, which are proportional to $\sqrt{nB}$~\cite{GusyninJPCM2006}. In contrast, WSMs exhibit a similar magneto-optical conductivity, but with a linear background that reveals the dispersion of LLs of WSMs along the direction of the applied magnetic field~\cite{AshbyPRB2013}.

Now we calculate the optical conductivity tensor in the vortex state of the generic WSC, which can be obtained from the Kubo formula~\cite{Ahn.NatureC2021,XuTianruiPRB2019,KamataniPRB2022,PapajPRB2022}. In the LL basis and in the clean limit, we have
\begin{eqnarray}
\sigma_{\alpha \alpha^\prime}(\omega)=- \tfrac{ie^2}{2\pi l_B^2} \sum_{nn^{\prime}} \int \tfrac{d{\bf k}}{(2\pi)^3} \tfrac{f(E_n)-f(E_{n^\prime})}{E_n-E_{n^\prime}} \times \tfrac{V_{\alpha}^{nn^\prime} V_{\alpha^\prime}^{n^\prime n}}{\omega+E_n -E_{n^\prime} + i\delta},
\end{eqnarray}
where $\alpha(\alpha')=\{x,y,z\}$, $d{\bf k}=dk_xdk_ydk_z$, and $f(\epsilon)=1/(1+e^{(\epsilon - \epsilon_f)/k_BT})$ is the Fermi-Dirac distribution function with $\epsilon_f=0$ for Bogoliubov quasiparticles and $k_B \equiv 1$ in our numerical calculations. The velocity matrix element $V_{\alpha}^{nn^\prime} = \langle \psi_{n{\bf k}} |V_{\alpha}| \psi_{n^\prime{\bf k}} \rangle$ with $| \psi_{n{\bf k}} \rangle$ being the eigenvector of the $n$-th LL
and the velocity operator given by~\cite{Ahn.NatureC2021,PapajPRB2022}
\begin{eqnarray}
V_{\alpha} ({\bf k}) =
\begin{pmatrix}
\partial_{k_\alpha} H^{\bf B}_0({\bf k}-e{\bf A}) & 0 \\
0 & \partial_{k_\alpha} {H_0^{\bf B}}^*(- {\bf k}-e{\bf A})
\end{pmatrix},
\end{eqnarray}
where $H^{\bf B}_0$ can be obtained from the BdG Hamiltonian processed through the singular gauge transformation.
\begin{figure}[ht]
  \centering
  \subfigure{\includegraphics[width=3.2in]{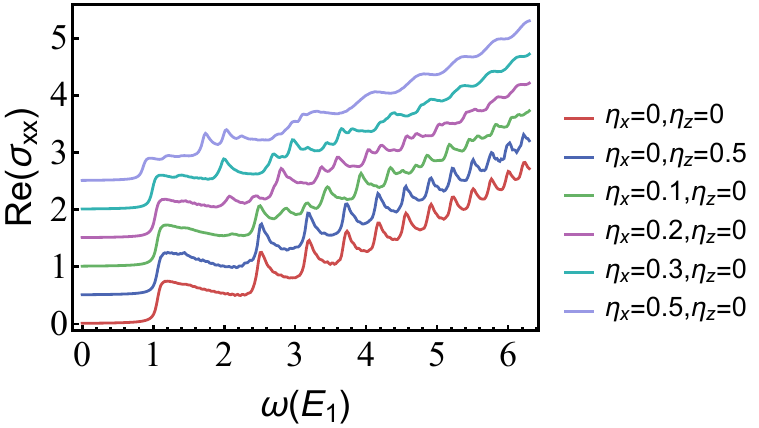}}
  \caption{The real part of the longitudinal optical conductivity (in units of $e^2/2\pi l_B$) in the mixed state of the generic WSC with tilted type-I BW cones, where $\lambda=0$, $l_B=14$, $\Delta_0=0.1$, $T=0.001$, and $\delta=0.01$. The curves have been displaced up for different values of $\boldsymbol{\eta}$. The red line denotes the $\boldsymbol{\eta}=0$ case.} \label{fig:MOCS}
\end{figure}

In Fig.~\ref{fig:MOCS}, we show the real part of the calculated longitudinal magneto-optical conductivity of the WSC with tilted type-I BW cones ($|\boldsymbol{\eta}|<1$). The magneto-optical conductivity for the untilted BW cones, represented by the red line, shows peaks only at photon frequencies $\omega/E_1=\sqrt{n}+\sqrt{n+1}$ for $n \ge 0$. This is because the primary contribution to the optical conductivity comes from the optical transitions between $E_{n,-}=-\sqrt{n}E_1$ and $E_{n\pm 1,+}=\sqrt{n\pm 1}E_1$ levels when the tilt $\boldsymbol{\eta}=0$. Additionally, the optical conductivity displays a linear background, which results from the dispersion of the LLs in the $k_z$ direction. When $\boldsymbol{\eta}\neq 0$, the tilt in the $k_z$ direction has little effect on the magneto-optical conductivity of the WSC, as indicated by the blue line. In contrast, the tilt $\eta_x$ shifts the conductivity peaks to lower optical frequencies due to the compression of LL spacings. Moreover, the usual dipolar selections $n\to n\pm 1$ are violated due to $\eta_x$, which leads to the emergence of new peaks arising from the wide $n\to m$ interband transitions in the optical conductivity. At low frequencies and small $\eta_x$, we can see well-pronounced lines corresponding to individual transitions, as depicted in Fig.~\ref{fig:MOCS}. Conversely, in the case of high frequencies and large $\eta_x$, the lines of multiple allowed transitions coalesce~\cite{JuditPRB2015}.

\begin{figure}[ht]
  \centering
  \subfigure{\includegraphics[width=3.2in]{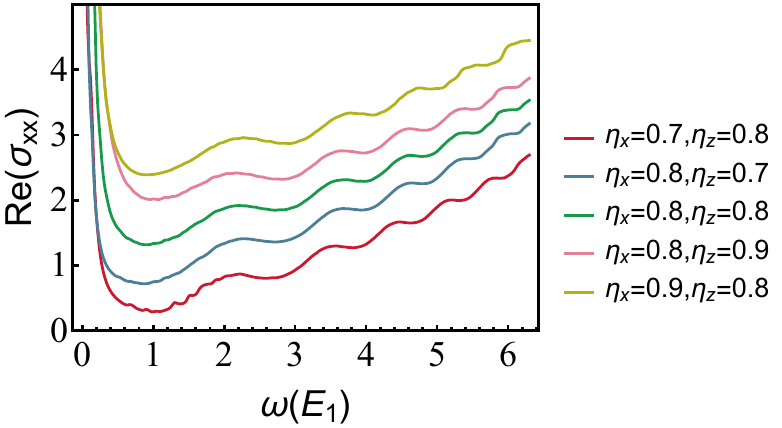}}
  \caption{The real part of the longitudinal optical conductivity (in units of $e^2/2\pi l_B$) in the mixed state of the generic WSC with type-II BW cones ($|\eta_x|<1$), where $\lambda=0$, $l_B=14$, $\Delta_0=0.1$, $T=0.001$, and $\delta=0.01$. The curves have been displaced up for different values of $\boldsymbol{\eta}$.} \label{fig:MOCII}
\end{figure}
Now let us consider the case $|\boldsymbol{\eta}|>1$, where we impose the restriction $|\eta_x|<1$ for LL quantization. In this case, the WSC model possesses type-II BW cones, which can exhibit special LLs that cross the quasiparticle Fermi level, as we can see in Fig.~\ref{fig:LLZf}. As a consequence, intraband optical transitions become feasible. Indeed, different from the $|\boldsymbol{\eta}|<1$ case, the optical conductivity in the mixed state of the WSC with type-II BW cones shows unique intraband peaks at low frequency, as shown in Fig.~\ref{fig:MOCII}. Such intraband conductivity peaks can serve as indicators to discern between type-I and type-II BW cones in WSCs. It should be noted that the conductivity peaks corresponding to individual transitions at higher frequency for the type-II BW cones are smeared out due to the larger $\eta_x$. At last, we want to illustrate that, as analyzed above, the tilt $\eta_x$ (in-plane electric field) may be eliminated by a Lorentz boost, leaving only a reduced magnetic field. Therefore, naively, the usual dipolar selection rule would not be violated in the optical conductivity. Actually, the novel conductivity peaks that arise from $n\to m$ interband transitions can be interpreted as the result of rotation symmetry breaking caused by the Lorentz boost back to the laboratory frame~\cite{LukosePRL2007,GoerbigEPL2009,JuditPRB2015}.

\section{Summary}\label{sec:Sum}
In this work, we proposed a WSC model that is engineered by alternately stacking WSM and $s$-wave superconductor layers. This model may exhibit tilted type-I and type-II BW nodes in the superconducting gap by tuning a tilt parameter $\boldsymbol{\eta}$. The type-II BW node contacts an electron-like and a hole-like quasiparticle pocket when $|\boldsymbol{\eta}|>1$. We investigated the quasiparticle spectrum in the mixed state of the WSC model. We found that, for type-I BW cones, tilted or not, the quasiparticle states in the mixed state of the WSC are always LLs. In contrast, for the type-II cones, the quasiparticle band structures depend on the angle between the direction of cone tilting and the magnetic field. When $|\boldsymbol{\eta}|>1$, and the projection $|\eta_{\bot}|$ in the plane perpendicular to the external magnetic field is less than $1$ (which is the Fermi velocity $v_F$ in the continuum model), LL quantization is still possible. Otherwise, the quasiparticles in the mixed state of the WSC behave as Bloch waves. In other words, only the tilt $\boldsymbol{\eta}_{\bot}$ can squeeze the LL spacings and induce the LLs collapse, while $\eta_z$ simply tilts the LLs in the $k_z$ direction.

We also studied the optical responses in the mixed state of the generic WSC. When $\boldsymbol{\eta}=0$, the longitudinal magneto-optical conductivity of the WSC model shows peaks only at optical frequencies $\omega \propto \sqrt{n} +\sqrt{n+1}$ for $n \ge 0$, with a linear background due to the dispersion of its LLs in the $k_z$ direction. This indicates that for the untilted BW cones, only the $n\to n\pm 1$ interband optical transitions are permitted, which is consistent with the usual dipolar selection rule in the magneto-optical conductivity of topological semimetals. When considering a nonzero tilt of BW cones in the $k_x$ direction, the usual dipolar selections are violated, giving rise to novel conductivity peaks that arise from the $n\to m$ interband transitions. Meanwhile, the tilt in the $k_z$ direction produces little impact on the magneto-optical conductivity of the WSC. When $|\boldsymbol{\eta}|>1$ ($|\eta_x|<1$), the BW cone tilts into a type-II cone, which exhibits unique intraband conductivity peaks at low frequency due to the presence of the LLs that cross the Fermi level. These intraband conductivity peaks provide a method to discern between type-I and type-II BW cones in WSCs.

\acknowledgments
This work was supported by Guangdong Basic and Applied Basic Research Foundation (Grant No. 2021B1515130007), Shenzhen Natural Science Fund (the Stable Support Plan Program 20220810130956001) and National Natural Science Foundation of China (Grant No. 12004442).

\normalem
\bibliography{WeylSC}
\bibliographystyle{apsrev4-2}

\end{document}